\documentclass{WileyMSP-template}
\usepackage{braket}
\usepackage{amssymb}

\begin{document}

\pagestyle{fancy}
\rhead{\includegraphics[width=2.5cm]{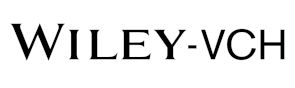}}

\title{Real-time capable CCD-based individual trapped-ion qubit measurement}

\maketitle


\author{Sebastian Halama}
\author{Timko Dubielzig}
\author{Niklas Orlowski}
\author{Celeste Torkzaban}
\author{Christian Ospelkaus}



\begin{affiliations}
S. Halama, T. Dubielzig, N. Orlowski, C. Torkzaban, C. Ospelkaus\\
Leibniz Universität Hannover\\
Institut für Quantenoptik\\
Welfengarten 1, 30167 Hannover\\
Germany\\
Email Address: halama@iqo.uni-hannover.de\\
\vspace{0.5cm}
C. Ospelkaus\\
Physikalisch-Technische Bundesanstalt\\
Bundesallee 100\\
38116 Braunschweig\\
Germany
\end{affiliations}


\keywords{surface-electrode ion trap, individual detection, EMCCD camera}

\begin{abstract}

Individual-qubit readout is a key ingredient for quantum simulation and quantum computation. To implement quantum error-correction protocols, readout must happen with a defined latency. In this paper the capability of an electron multiplying (EMCCD) camera with a real-time processing capable output is demonstrated to determine the quantum state of single $^{9}\mathrm{Be}^{+}$ ions and the required timing sequences are explored. The results are found to be comparable to the commonly used detection method based on photomultiplier tubes. We report experiments on the individual detection of $^{9}\mathrm{Be}^{+}$ qubit states undergoing coherent excitation to characterize the readout performance. Sources of error and the amount of crosstalk in the detection system are discussed. Error rates due to known problems in the state preparation and measurement processes were determined to be approximately $0.5\,\%$. 

\end{abstract}

\section{Introduction}
Trapped ions are a promising candidate~\cite{turchette_deterministic_1998} for the developement of quantum simulators~\cite{blatt_quantum_2012} and quantum computers~\cite{pino_demonstration_2021} with a large number of qubits that can outperform the simulation and computation capabilities of classical computers. A critical step for quantum computers and simulators is the ability to read out the state of qubits through state-dependent fluorescence detection. Pioneering experiments relied on global fluorescence detection using a photon-counting detector~\cite{sackett_experimental_2000}. The implementation of arbitrary algorithms requires the capability to read out individual qubits while leaving others unaffected. To implement error-correction protocols for example, the readout needs to be completed in a defined time, which is short compared to relevant decoherence time scales. In scalable trap arrays, e.g. based on surface-electrode trap technology~\cite{chiaverini_surface-electrode_2005,seidelin_microfabricated_2006}, dedicated detection zones can be implemented where ions are locally illuminated near trap-integrated detectors, and individual ions can be shuttled into such dedicated readout registers. Several approaches have been made to demonstrate scalable individual qubit detection in ion traps. Towards this end, photodiodes~\cite{eltony_transparent_2013,setzer_fluorescence_2021}, superconducting nanowire single-photon detectors (SNSPDs)~\cite{todaro_state_2021}, and fibers~\cite{vandevender_efficient_2010} have been integrated into trap chips to perform individual qubit detection. On the other hand, it can be desirable to implement individual-ion detection in longer ion chains or in ion-trap arrays for quantum simulations~\cite{porras_bose-einstein_2004,porras_effective_2004,schmied_quantum_2011,lemmer_trapped-ion_2018}. Spatially resolving devices such as multi-channel photomultiplier tubes (PMTs) \cite{debnath_demonstration_2016} or electron multiplying charge-coupled device (EMCCD) \cite{burrell_scalable_2010} cameras have been used for this purpose. Multi-channel PMTs provide an individual-ion readout capability and provide the photon counts in realtime, but are limited in flexibility because of their fixed detector sizes and positions. An EMCCD camera, on the other hand, comes with a slight disadvantage in readout speed, but is highly flexible since it can measure the fluorescence of individual ions at arbitrary positions in the trap.

Here we investigate the capabilities of a real-time capable EMCCD camera for individual qubit detection and compare the results with similar measurements taken with a PMT. We first present experiments of Rabi oscillations and the resulting photon count distributions for a single ion. We then characterize the camera by simultaneously reading out the states of two ions and demonstrate its ability to individually detect the qubit states after individual-ion coherent state manipulation.

\section{Experimental overview}
The experimental setup was discussed in detail in \cite{dubielzig_ultra-low-vibration_2021}. The setup has been built to explore control of the internal and motional states of $^{9}\mathrm{Be}^{+}$ ions using a microwave near-field approach~\cite{ospelkaus_trapped-ion_2008}, which was demonstrated to work in a room-temperature experiment~\cite{hahn_integrated_2019} using a trap similar to the one in the present cryogenic system. The ion trap is a surface-electrode Paul trap, which consists of DC and RF electrodes used to generate a three-dimensional trapping potential, together with three integrated microwave control conductors. Two of these can be used to drive carrier transitions, while the third has a special shape that enables it to generate a localized oscillating magnetic field gradient at the ions' position for motional sideband transitions and multi-qubit quantum logic~\cite{ospelkaus_trapped-ion_2008} while simultaneously having a low background field to prevent off-resonant carrier coupling~\cite{carsjens_surface-electrode_2014,wahnschaffe_single-ion_2017}. To load the trap with Beryllium ions, we use ablation loading with $1064\,\mathrm{nm}$ laser pulses impinging on a Beryllium wire~\cite{dubielzig_ultra-low-vibration_2021,dubielzig_ultra-low_2020}. Some of the atoms emitted travel through the trap center and are ionized by a $235\,\mathrm{nm}$ laser beam in a resonant two-photon process~\cite{lo_all-solid-state_2014}. Ions are then Doppler cooled using a $313\,\mathrm{nm}$ laser beam~\cite{wilson_750-mw_2011}.

We use an ultralow vibration closed-cycle cryostat to operate the trap at a temperature of $5\,\mathrm{K}$~\cite{dubielzig_ultra-low-vibration_2021}, which results in an excellent vacuum that enables us to keep ions trapped for several days at a time.

The axial confinement in the trap is realized by applying voltages between $-10\,\mathrm{V}$ and $+10\,\mathrm{V}$, which are generated by an arbitrary waveform generator based on a field-programmable gate array (FPGA)~\cite{bowler_arbitrary_2013} and applied to the 10 DC-electrodes. The radial confinement is ensured by applying an RF frequency of $99.365\,\mathrm{MHz}$ to a cryogenic helical resonator~\cite{dubielzig_ultra-low-vibration_2021,dubielzig_ultra-low_2020} matched to the $50\,\mathrm{\Omega}$ feed line. We find trap frequencies of $\omega_{\mathrm{ax}}=2\pi\cdot 1.081\,\mathrm{MHz}$, $\omega_{\mathrm{r1}}=2\pi\cdot 11.32\,\mathrm{MHz}$ and $\omega_{\mathrm{r2}}=2\pi\cdot 11.53\,\mathrm{MHz}$.

\subsection{Imaging system}
A cryogenic Schwarzschild objective with a numerical aperture (NA) of approximately 0.34 and magnification of 40 is positioned at a working distance of $8\,\mathrm{mm}$ from the trap and collects fluorescence photons emitted by the ions~\cite{dubielzig_ultra-low_2020}. The light from the ions can be focused either onto a PMT or an EMCCD camera using two lenses and a flip mirror located in a lightproof box under the optical table. In contrast to a PMT, the camera can spatially resolve the source of the fluorescence light and therefore enables individual-ion state detection, which is one key requirement for future quantum simulators and quantum computers~\cite{divincenzo_physical_2000}.

\subsection{EMCCD camera}
The measurements presented in this paper were performed with an Andor iXon Ultra 888 UVB, which has several important features required for our experiments. This camera can detect UV light at $313\,\mathrm{nm}$ with a relatively high quantum efficiency of $\approx33\,\%$~\cite{andor_ixon_2019} comparable to the PMT (Hamamatsu H10682-210)~\cite{hamamatsu_photonics_kk_photon_2016}; this high quantum efficiency is essential in order to produce a sufficient amount of counts to be able to distinguish betweeen the dark state and the bright state.

Hardware binning enables us to combine multiple pixels into one binning area per ion, which can be seen as one single big pixel. When choosing a binning area, it is necessary to strike a balance between a large enough area to detect enough of the photons scattered by the ion, but not so large that it would detect the light scattered by a neighboring ion; this crosstalk can lead to measurement errors that reduce the fidelity of the individual-ion state detection. Usually we set binning areas of approximately 30\,x\,30 pixels around each of the ions and leave a gap of five to ten pixels between the binning areas of neighbouring ions (see section \ref{sec:crosstalk}).

The electron multiplying (EM) amplifier allows us to combine the single photon sensitivity of a PMT with the spatial resolution of a CCD camera, and the hardware binning produces significantly less electronic readout noise than software binning would since the noisy readout process is performed only once per ion. The camera settings and data acquisition can be controlled by the Andor Solis software package, but a major feature of the approach presented here is that the camera possesses a Camera Link real-time output which can be used to process the data in real-time e.g. for quantum error correction. The ARTIQ real-time environment already features an expansion module for this bus, and in preliminary experiments we have been able to read out synthetic data in real-time. For the present manuscript, as the experiment is still controlled by the system described in~\cite{langer_high_2006}, we extract the data from the camera's host computer after a measurement series is completed. 

The camera's charge-coupled device (CCD) chip can be thermo-electrically cooled down to around $-100\,\mathrm{^{\circ}C}$ when in idle state. Lower temperatures lead to a better discriminability between the dark state and the bright state, which we trace back to reduced electronic noise and also different internal signal processing of the camera's firmware, that we observed to depend on the set CCD temperature. On the other hand, lower temperatures allow only for a smaller number of readout operations per second, because every shift and readout operation introduces heat in the chip and the overall cooling power is limited. We find a reasonable compromise at a temperature of $-35\,\mathrm{^{\circ}C}$, where we can achieve up to 200 readout operations per second and a ``state preparation and measurement" (SPAM) error comparable to the error we obtained when detecting the signals with a PMT.

We require exposure times of $400\,\mathrm{\mu s}$, too short for the camera's built-in mechanical shutter, which is therefore set to always open. The resulting permanent illumination leads to accumulated charges on the CCD sensor; most of these are generated during the Doppler cooling sequence, during which fluorescence light from the cooling process illuminates the camera. We clean the sensor by triggering a readout sequence in between Doppler cooling and the actual readout event (see figure \ref{fig:brightdarkpulsesequence}); the image taken this way contains no useful information and is discarded during processing. After cleaning the sensor, we shine in the detection laser for $400\,\mathrm{\mu s}$ and read out the CCD sensor a second time. This means that with 200 images taken per second, only 100 experiments can be run in this time. In a scenario with only one Doppler cooling sequence followed by multiple qubit manipulations and intermediate readout operations, the readout rate would asymptotically increase to $200\,\mathrm{s}^{-1}$. 

\section{Single-qubit performance}

\subsection{Experimental sequence}
\begin{figure}
	\centering
	\includegraphics[]{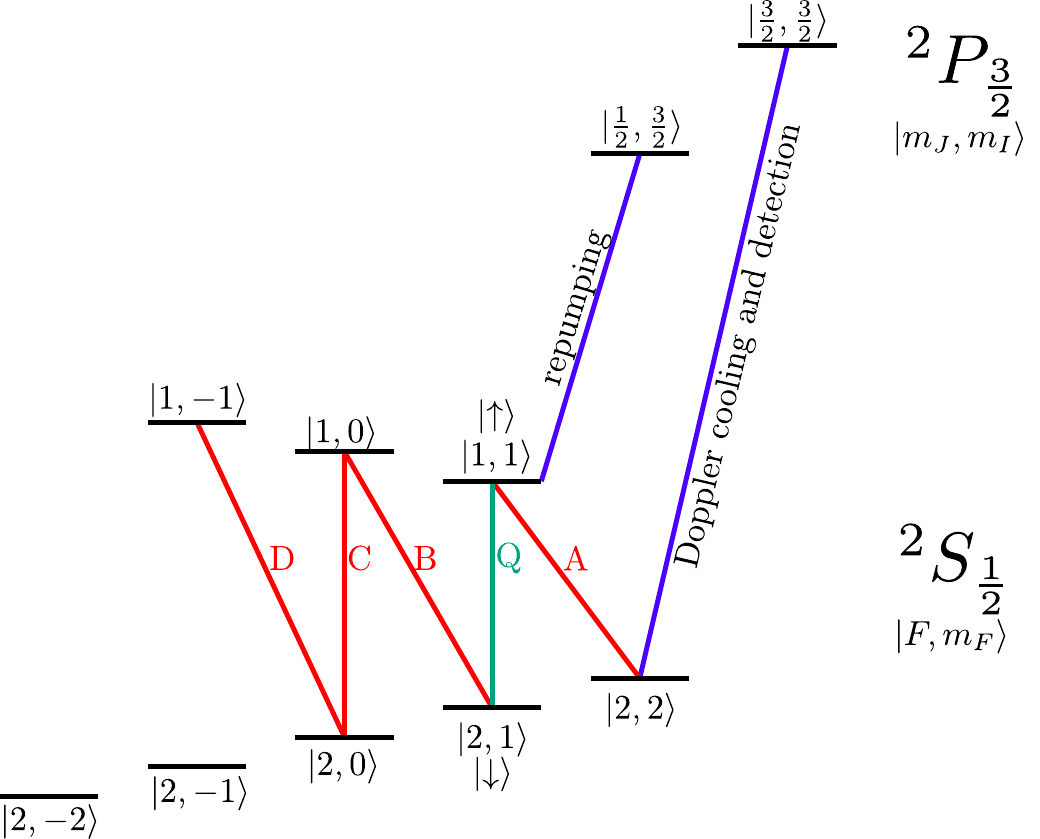}
	\caption{Relevant transitions in $^{9}\mathrm{Be}^{+}$ ions. The green line indicates the qubit transition with a frequency of $1082.55\,\mathrm{MHz}$. This transition and the other transitions indicated as red lines can be driven with microwave pulses. The blue lines indicate the cooling and detection transition as well as the repumping transition, which are driven with laser beams at a wavelenght of $\approx313\,\mathrm{nm}$.}
	\label{fig:energylevels}
\end{figure}

\begin{figure}
	\includegraphics[width=\linewidth]{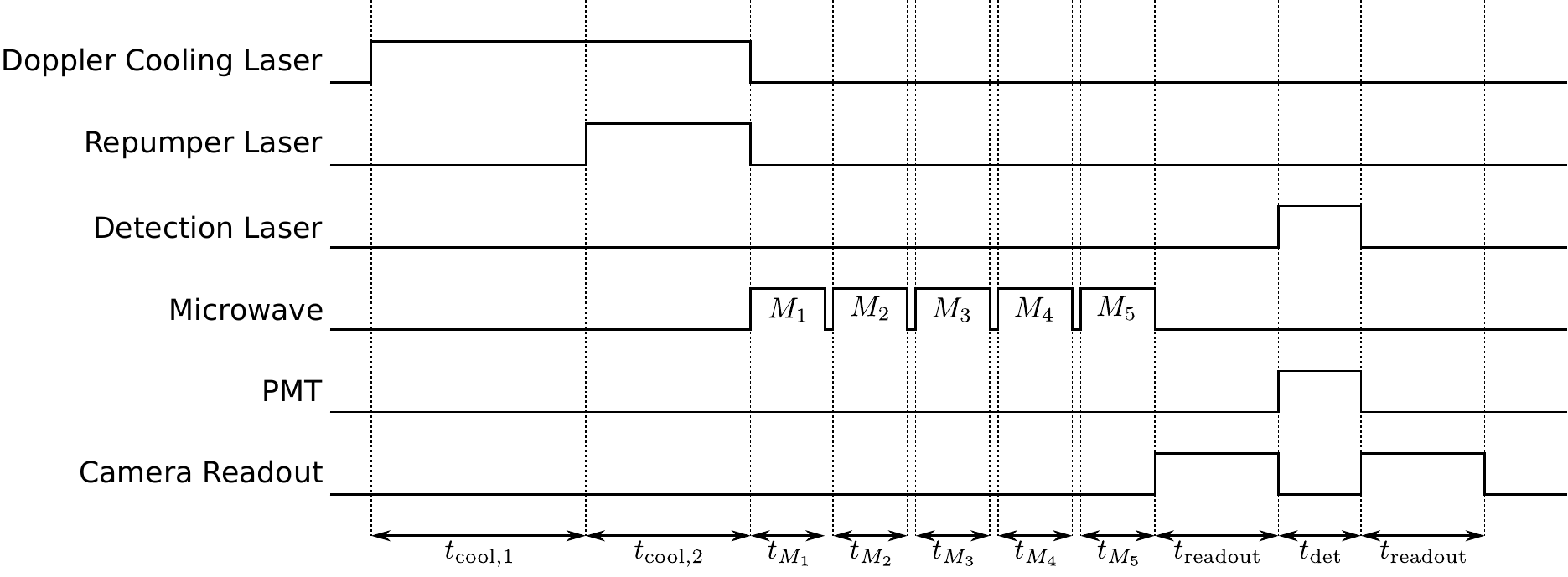}
	\caption{Simplified pulse sequence for the experiments presented in this paper.
		Time runs from left to right.
		Pulse durations are not to scale.
		$t_{\textrm{cool},1} = 5\,\mathrm{ms}$,
		$t_{\textrm{cool},2} = 1\,\mathrm{ms}$,
		$t_{M_i}$ dependent on experiment,
		$t_{\textrm{readout}}\approx 2\,\mathrm{ms}$,
		$t_{\textrm{det}}=400\,\mathrm{\mu s}$.}
	\label{fig:brightdarkpulsesequence}
\end{figure}

Here we give an overview of the experiments we have performed using a single $^{9}\textrm{Be}^{+}$ ion. With an external magnetic field of $B_0 = 22.3\,\textrm{mT}$ at the ions' position, the qubit transition frequency between the two hyperfine states $\ket{1,1} \equiv \ket{\uparrow}$ and $\ket{2,1} \equiv \ket{\downarrow}$ is magnetic-field-independent to first order. Here the notation $\ket{F,m_F}$ is used to describe states in the $^{2}S_{1/2}$ manifold and $\ket{m_J,m_I}$ is used to describe states in the ${^{2}P_{3/2}}$ manifold. An experiment starts with optical pumping to the $\ket{2,2}$ state. A Doppler cooling laser drives the closed-cycle transition between $^{2}S_{1/2}\ket{2,2}$ and ${^{2}P_{3/2}}\ket{\frac{3}{2},\frac{3}{2}}$ for $6\,\mathrm{ms}$. During the final $1\,\mathrm{ms}$, we also shine in a repumper laser beam on the $^{2}S_{1/2}\ket{1,1} \rightarrow {^{2}P_{3/2}}\ket{\frac{1}{2},\frac{3}{2}}$ transition (see figures \ref{fig:energylevels} and \ref{fig:brightdarkpulsesequence}). Since this is not a closed-cycle transition, there is a chance that the ion decays into the $\ket{2,1}$ or $\ket{2,2}$ state. This helps to depopulate the $F=1$ states and therefore increases the resulting population in the $\ket{2,2}$ state. Starting here we can then transfer the population to any state (or superposition of states) in the $^{2}S_{1/2}$ manifold using microwave pulses.

We apply a microwave $\pi$-pulse on transition A and thereby transfer the population to the $\ket{1,1}$ state to initialize the qubit. We can now perform operations on the qubit transition $\ket{1,1}\leftrightarrow\ket{2,1}$ labeled Q. To detect the qubit state, we first shine in a $\pi$-pulse resonant with transition A to transfer the population in the $\ket{1,1}$ state to the $\ket{2,2}$ state. After that, we apply a sequence of three microwave $\pi$-pulses resonant with transitions B, C and D to transfer the population in $\ket{2,1}$ to $\ket{1,-1}$. The Doppler cooling laser is also used as the detection laser by shifting its frequency approximately $10\,\mathrm{MHz}$ towards the resonance. The population in $\ket{2,2}$, the bright state, scatters the highest number of photons from the detection laser beam, which are then collected by the Schwarzschild objective and focussed onto either the PMT or the EMCCD camera. The population in $\ket{1,-1}$, the dark state, scatters the least amount of light, because the detection laser has the highest detuning from any state in the $P_{3/2}$ manifold when starting in this state, leading to a minimal number of detected photons.

To be able to compare the measurements performed with the PMT and the camera, we designed the detection sequence in such a way that it works for both detection methods independent of the actual detector used. When using the EMCCD camera for detection, one has to consider that photons scattered during optical pumping may have already been detected by the CCD and produced charges in the sensor. To clean the CCD sensor before the detection sequence is performed, we trigger a readout after the final microwave pulse. After the $2\,\mathrm{ms}$ needed to complete the readout of the CCD with the currently used settings, we then shine in the detection laser for $400\,\mathrm{\mu s}$. This is also the timeslot during which the PMT events are counted by the experiment's control system.
We perform a second readout of the camera's CCD sensor after switching off the detection laser, which again takes $2\,\mathrm{ms}$ to complete.

Both PMT and camera detection procedures yield a number of counts that we use to measure the qubit's state. There are several error sources that are all condensed in the SPAM error. First of all, an imperfect circular polarization of the laser beams can either lead to an imperfect preparation of the $\ket{2,2}$ state or it can cause the ion to leave the closed-cycle transition used for state detection.
In both cases the ion is excited to undesired states in the $P_{3/2}$ manifold, which then decay to a state different from the $\ket{2,2}$ state. The second and most likely smallest error source is an imperfect transfer between the qubit states and the dark and bright states, which could happen when using the incorrect microwave frequency or pulse duration.
The third error source is a finite chance for the detection laser beam to excite the population in the $\ket{1,-1}$ state, which then leads to so-called optical depumping to the bright state, resulting in a high count rate despite the ion initially being in the dark state.

\subsection{Single-ion qubit coherent rotations}
\begin{figure}
	\centering
	\includegraphics[]{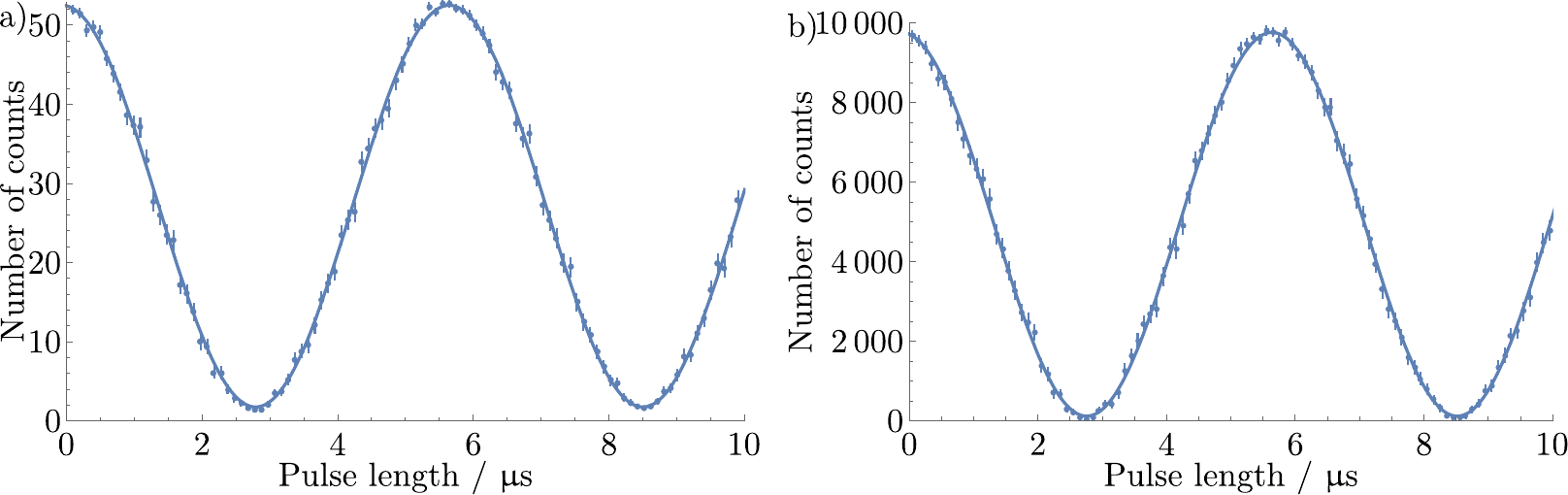}
	\caption{Rabi oscillations on the qubit transition.
		a) PMT detection, b) camera detection.
		Dots represent the average number of counts measured in 500 experiments and the error bars indicate the standard deviation.
		For the camera detection, the average baseline is subtracted.
		The solid line is a sinusoidal fit to the data.		
	}
	\label{fig:1ionflopping}
\end{figure}

In this section we present the results for coherent state manipulation of the qubit transition. The detection of the qubit state was done one time with a PMT and another time with the EMCCD camera, and the corresponding data is shown in figure \ref{fig:1ionflopping}.

For this measurement the ion is prepared in the $\ket{\uparrow}$ state followed by a microwave pulse of varying length in steps of $0.1\,\mathrm{\mu s}$. For every microwave pulse length, the experiment was repeated 500 times and the number of counts was averaged over these experiments. Both measurements can be used to extract the $\pi$-time of the qubit transition.

For an accurate determination of the relative amount of background counts for these curves, we have to account for the camera's non-zero baseline counts. We closed the mechanical shutter of the camera and triggered the readout process in the same way we do when measuring the ion's fluorescence, and the baseline was found to be 925.0(2.8). The baseline countrate of the PMT does not have to be taken into account since it is basically zero.

A side-by-side comparison of the two population oscillation curves is shown in figure \ref{fig:1ionflopping}. 
The relative amount of background counts is calculated by $\beta=\frac{\textrm{Min}}{\textrm{Max}-\textrm{Min}}$ and we use the maxima and minima of the fitted curves to calculate $\beta$.
For the PMT detection, we find $\beta_{\textrm{PMT}}=3.46(30)\,\%$, while for the camera detection we achieve $\beta_{\textrm{Cam}}=1.24(27)\,\%$. Because only a small part of the CCD is read out, the detection with the camera is much less sensitive to stray light compared to using a PMT and therefore has a lower background. To reach similar values with a PMT, one would have to carefully place apertures to block the stray light from reaching the PMT.

\subsection{Single-ion detection}
\begin{figure}
	\centering
	\includegraphics[]{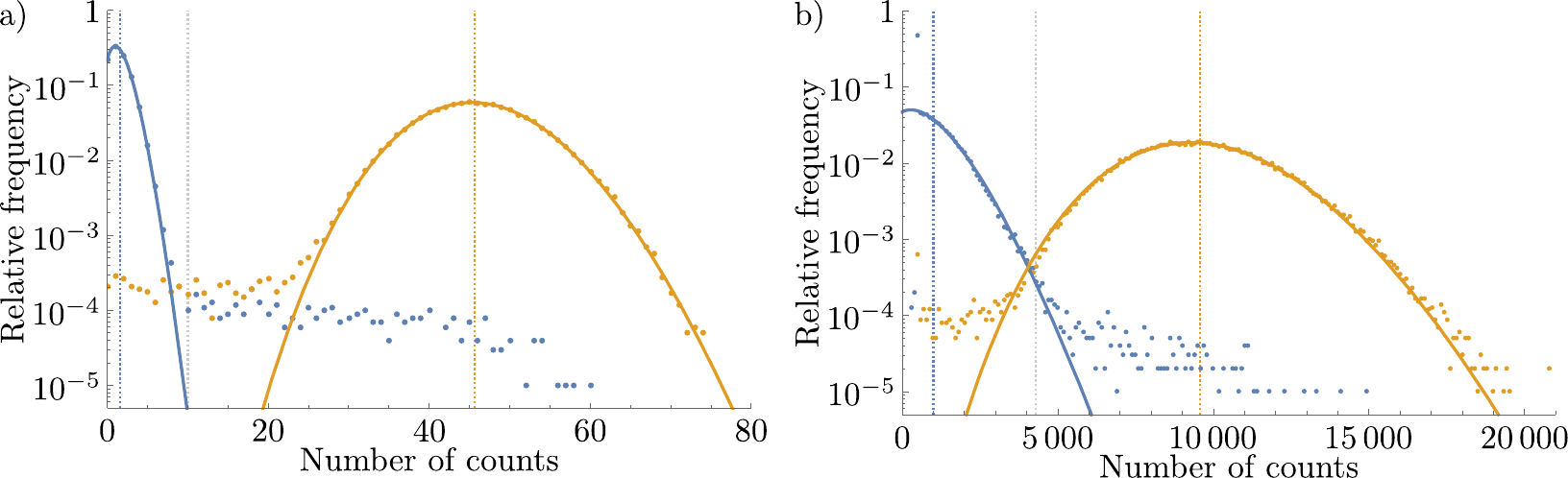}
	\caption{Detection histograms for $10^5$ experiments for each preparation. 
	a) PMT detection, b) camera detection (100 count values on the horizontal axis are binned together for clarity).
	blue: preparation in the dark state, orange: preparation in the bright state.
	Points: experimental data, solid lines: Poissonian fit (datapoint at 500 and below were removed for fitting camera data), dashed lines: average number of counts for dark and bright preparation. 
	The grey dashed line indicates the threshold value to decide between dark and bright.
	}
	\label{fig:1ionhistograms}
\end{figure}

In this section we present the resulting photon count distribution when detecting the qubit state after initializing it in either the $\ket{\uparrow}$ state or the $\ket{\downarrow}$ state and subsequently transferring it to the bright state or the dark state. A histogram representation of the measurement is shown in figure \ref{fig:1ionhistograms}.

When we repeat each experiment $10^5$ times, we can determine from the average number of counts whether the ion was most likely in the bright state or in the dark state. For the preparation in the bright state we obtain on average $45.642(23)$ counts on the PMT and $9568(7)$ counts on the camera, while for the preparation in the dark state we detect on average $1.635(7)$ counts on the PMT and $998.9(2.4)$ on the camera.

There are some clearly visible differences between the two histograms. First of all the absolute number of counts for the camera detection is a lot higher than for the PMT detection. For a PMT measurement the number of counts is proportional to the number of incident photons on the detector when neglecting dead time effects. In case of the camera detection the incident light will generate charges in the individual pixels of the CCD chip; we use hardware binning of 30 x 30 pixels to collect the light from a larger detector area compared to individual pixel readout. The binned charges are shifted to the EM-amplifier and a bias voltage is applied before entering the analog-digital-converter. While the bias voltage leads to a baseline of 500 counts which is set by the camera firmware, the amplifier leads to higher counts compared to a PMT. There is a noticeable outliner at exactly 500 counts during the camera detection process for very low light levels. 
This number is exactly the baseline value and the presence of this spike is caused by the EMCCD camera's firmware not being optimized for the mode of operation we are using.
The height of this point at 500 counts depends on the used camera settings, but it has no effect on the following SPAM-error estimation, as long as the threshold value for discriminating between the dark state and the bright state is above 500 counts.

It is also evident that the camera's detection process imprints some noise on the histograms, which broadens the peaks. We presume that the noise is caused by the shift operations in the CCD or by the readout amplifier. For the PMT detection, there is a valley between the peaks for dark and bright preparation procedures, where both histograms are almost flat. In contrast to that, in the camera's data, the falling and rising slopes of the histograms corresponding to the dark and bright preparation procedures intersect. 

\begin{figure}
	\centering
	\includegraphics[]{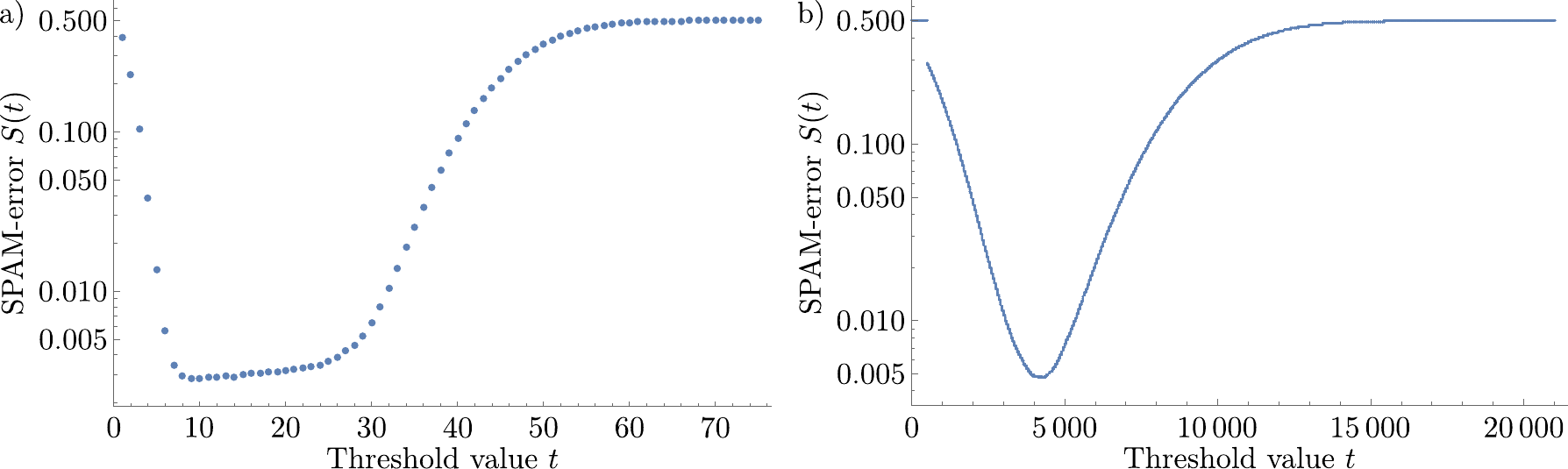}
	\caption{Resulting SPAM error depending on the threshold value to decide between the dark state and the bright state.
	a) PMT detection, b) camera detection.	
	The jump in b) is due to the peak at a number of counts of 500 in figure \ref{fig:1ionhistograms}.}
	\label{fig:spamplots}
\end{figure}

To implement quantum algorithms, it is crucial to detect the state of a qubit with a single detection event. Therefore we have to rely on the counts detected during a single measurement. Since the two histograms for the bright and dark states overlap, there will be an error during the detection process. From the data shown in figure \ref{fig:1ionhistograms}, we determined a threshold value for the detection to determine if the ion was in the dark state or in the bright state. Every measurement with counts below that value is interpreted as corresponding to the dark state, while measurements with counts equal to or above the threshold are interpreted as corresponding to the bright state. The threshold was experimentally set so that the SPAM error is minimal (see figure \ref{fig:spamplots}). For any threshold value $t\in\mathbb{N}$, we calculate the resulting SPAM error $S(t)$ by summing over all wrongly assigned detection events and dividing by the total number of detection events:
\[S(t)=\left(\sum_{i=0}^{t-1}N_{B}(i) + \sum_{i=t}^{\infty}N_{D}(i)\right)/\left(\sum_{i=0}^{\infty}N_{B}(i) + N_{D}(i)\right)\]
Here $N_{B/D}(i)$ is the number of times that we measure $i$ counts for bright/dark state preparation.
Note that with this method we cannot distinguish between an error during preparation and during detection, but we assume that the error contribution during the preparation process is much smaller due to the presence of the repumper laser.
Since the experiment was repeated an equal amount of times with an ion prepared in the bright state and in the dark state, the SPAM error converges to $50\,\%$ for very low and for very high threshold values. We choose the threshold values that yield the lowest SPAM error, which are 10 for the PMT and 4284 for the camera. Here the SPAM error is $\approx0.3\,\%$ when using the PMT and $\approx0.5\,\%$ when using the camera.

A potential source of error in our experiment could be imperfect $\sigma^{+}$ polarization of cooling and detection laser beams, which leads to so-called bright-to-dark depumping~\cite{hahn_two-qubit_2019,lo_creation_2015}. During detection, the population in the bright state $\ket{2,2}$ can be pumped to a different state that yields a lower number of counts, most likely the $\ket{1,1}$ state.
Although we use a polarizer and waveplates to set the correct polarization before the beam enters the vacuum chamber in which the ion trap is located, there could be spatially inhomogeneous dichroic effects of the cryogenic viewports on the polarization of the cooling and detection laser beams, which we cannot compensate from the outside.
Another source of error is off-resonant excitation of the dark state during the detection process.
This dark-to-bright depumping leads to a high number of detected photons despite the ion being initially in the dark state.
Both of the mentioned error sources lead to a deviation from the ideal poissonian shape of the histograms (c.f. figure \ref{fig:1ionhistograms}) and therefore introduce an error.

\section{Two-qubit single-shot performance}

\subsection{Experimental sequence}
We performed three different experiments on two ions and compare the results obtained with the PMT and the camera. In the first experiment the ions were both prepared in the $\ket{\uparrow}$ state followed by the detection sequence, in the second experiment both ions were prepared in the $\ket{\downarrow}$ state followed by the detection sequence, and in the third experiment the ions were prepared in the superposition state $\frac{1}{\sqrt{2}}(\ket{\uparrow}+\ket{\downarrow})$ followed by the detection sequence. This experiment was repeated $10^5$ times for each detection method: the PMT, the camera with a single binning area, and the camera with individual binning areas for each ion. 

\subsection{Two-qubit individual state readout}

\begin{figure}
	\centering
	\includegraphics[]{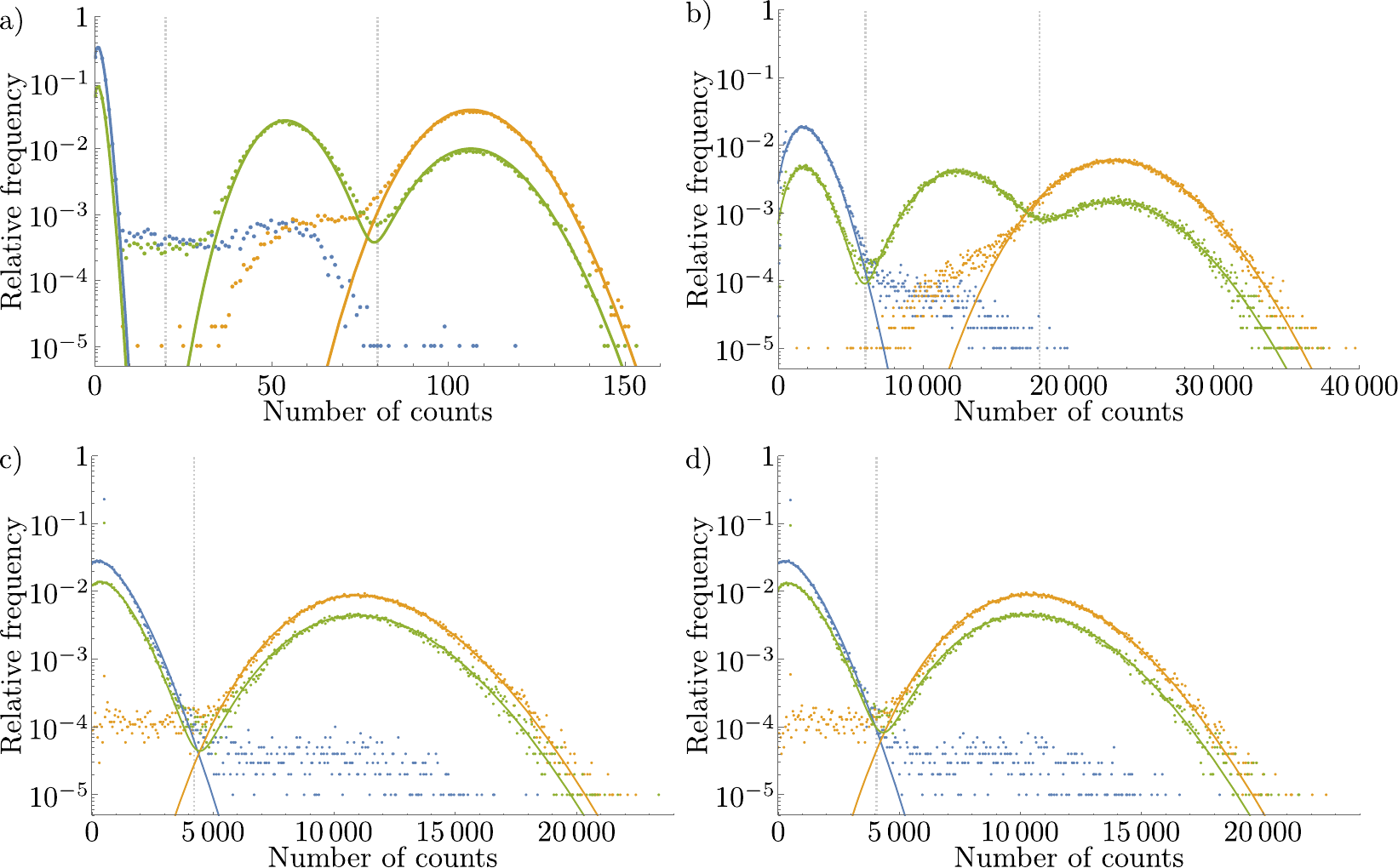}
	\caption{Histogram representation for two-ion detection. Points: experimental data, lines: Poissonian fit (datapoint at 500 was removed for fitting camera data). blue: preparation in the dark state, orange: preparation in the bright state, green: preparation in the superposition state.
	a) PMT detection, b) camera detection with one large binning area for both ions, c) camera detection with binning area for ion 1, d) camera detection with binning area for ion 2.
Vertical lines indicate the threshold values to decide between 0, 1 or 2 ions bright (a \& b) respectively dark state and bright state (c \& d)}
	\label{fig:2iondarkbrightpiHalfHistograms}
\end{figure}

The results when detecting the ions' states with the PMT are shown in figure \ref{fig:2iondarkbrightpiHalfHistograms}a. When looking at the fitted line, it is evident that for preparation in the dark state and the bright state, the respective histograms do not follow purely poissonian shapes (indicated by the fitted lines), but instead tend towards a flat line, indicating that one of the ions is pumped from dark to bright or vice versa, before decreasing further down towards zero. This effect is similar to the flat line for single-ion detection (see figure \ref{fig:1ionhistograms}). Here we observe twice the effect, since there are two ions in the trap. The vertical lines at 20 and at 80 counts indicate the threshold values to decide between 0, 1 or 2 ions bright for a single-shot detection. In table \ref{tab:TwoIonsPMTHist}, the distribution for the number of ions interpreted to be bright for a given preparation scheme is shown for PMT detection, and for camera detection with the threshold values set to 6000 and 18000 counts within a single binning area. The error $E_P$ for a given preparation scheme $P$ stated in the table is calculated in the following way:
\[E_P = \frac{1}{2}\sum_{i=0}^{2}\left|M(i,P)-T(i,P)\right|,\]
where $P\in\{D,S,B\}$ corresponds to preparation in the dark state (D), the superposition state (S) or the bright state (B), $M(i,P)$ is the number of times $i$ ions were measured to be bright for preparation scheme $P$, and $T(i,P)$ is the theoretically predicted value to find $i$ ions bright for preparation scheme $P$ neglecting any depumping effects. Ideally we would always find 0 ions bright following preparation in the dark state and 2 ions bright following preparation in the bright state. In the case that ions are prepared in a superposition state, it is expected to find 0 and 2 ions bright in $25\,\%$ of the experiments and 1 ion bright in $50\,\%$ of the experiments. The factor $\frac{1}{2}$ is a normalisation factor to prevent errors from being counted twice.

Both systems have a larger error than during single-ion experiments, where the ion was prepared in either the bright state or the dark state. In case of preparation in the superposition state, the individual errors somehow compensate each other, resulting in an error rate comparable to the SPAM error from single-ion experiments. We notice that the error for ions in the dark state is significantly lower for detection with the camera compared to the PMT, but drastically higher for ions in the bright state. We presume that the detection of ions in the dark state benefits from the nonlinear detector characteristic of the camera at low light level and that is also responsible for the peak at 500 counts. One can determine by eye that the blue datapoints decrease faster towards zero right of the first threshold value in figure \ref{fig:2iondarkbrightpiHalfHistograms}b than in figure \ref{fig:2iondarkbrightpiHalfHistograms}a. For ions in the bright state, the histogram is broadened due to noise when we detect with the camera as it was already observed for the experiments with one ion. Here both the peaks for one ion bright and two ions bright are broadened which leads to the increased overlap compared to the detection with the PMT. Although these noise processes are absent when detecting two ions with the PMT, the error rate is higher than for the single-ion case since the overlap of the histograms of one ion bright and two ions bright always has an effect on the error that decreases with the average number of counts detected for a single ion.

\begin{table}
	\centering
\begin{tabular}{|c|r|r|r|r|}
	\hline
	\textbf{Preparation} & \textbf{0 bright} & \textbf{1 bright} & \textbf{2 bright} & \textbf{error}\\
	\hline
	\multicolumn{5}{|c|}{Detection with PMT} \\
	\hline
	\textbf{dark} & 97\,642 & 2\,344 & 14 & $2.358\,\%$\\
	\hline
	\textbf{superposition} & 25\,469 & 50\,219 & 24\,312 & $0.688\,\%$\\
	\hline
	\textbf{bright} & 5 & 2\,769 & 97\,226 & $2.774\,\%$\\
	\hline
	\multicolumn{5}{|c|}{Detection with EMCCD camera} \\
	\hline
	\textbf{dark} & 98\,708 & 1\,285 & 7 & $1.292\,\%$\\
	\hline
	\textbf{superposition} & 25\,668 & 50\,144 & 24\,188 & $0.812\,\%$\\
	\hline
	\textbf{bright} & 10 & 6\,294 & 93\,696 & $6.304\,\%$\\
	\hline
\end{tabular}
\caption{\label{tab:TwoIonsPMTHist}Abundance of detecting 0, 1 or 2 ions bright using the threshold method for two ions prepared in the dark state, bright state or a superposition of these states. The experiment was repeated $10^5$ times for each preparation. Top: Detection with PMT, bottom: Detection with EMCCD camera using a single binning area.}
\end{table}

For the measurements shown in figure \ref{fig:2iondarkbrightpiHalfHistograms}c and \ref{fig:2iondarkbrightpiHalfHistograms}d, we detected the states of the ions individually by configuring individual binning areas for each ion. Here the threshold value is used to decide for each individual ion if it was in the bright state or in the dark state. The error is equivalent to the SPAM error for single-ion detection: with a threshold value of 4211 for the first ion, we achieve a SPAM error of $0.56\,\%$ and for the second ion with a threshold value of 4070, we achieve a SPAM error of $0.58\,\%$. These errors are comparable to the SPAM error during single-ion detection and are significantly lower than in the previously discussed experiments in which the common fluorescence light of both ions was detected.

With the individual detection capability, we can further investigate the $50\,\%$ propability for each of the ions to be detected in the bright state. Ideally this should decompose into $25\,\%$ for only the first ion to be bright and  $25\,\%$ for only the second ion to be bright. The results for this individual detection are shown in table \ref{tab:IndividualDetection}. Using the values from this table and the mapping ``dark"$\leftrightarrow\ket{\downarrow}$ and ``bright"$\leftrightarrow\ket{\uparrow}$, we can calculate that the propability for the first ion to be found in the $\ket{\downarrow}$ state is $P_1(\downarrow)=P(\downarrow\downarrow)+P(\downarrow\uparrow)=50.7\,\%$ and therefore significantly higher than for the second ion, which is $P_2(\downarrow)=P(\downarrow\downarrow)+P(\uparrow\downarrow)=50.1\,\%$. This is unlikely to be caused by measurement errors, but rather by other effects such as different Rabi rates for the two ions (see section \ref{sec:twoqubitmultishotperformance}).

\begin{table}
	\centering
	\begin{tabular}{|c|c|c|}
		\hline
		\multicolumn{2}{|c|}{\textbf{detected state}} & \\
		\cline{1-2}
		\textbf{Ion 1} & \textbf{Ion 2} & \textbf{relative abundance}\\
		\hline
		dark & dark & $25.4\,\%$ \\
		\hline
		dark & bright & $25.3\,\%$ \\
		\hline
		bright & dark & $24.7\,\%$ \\
		\hline
		bright & bright & $24.5\,\%$ \\
		\hline
	\end{tabular}
	\caption{\label{tab:IndividualDetection}Individual detection of two ions after a global $\pi/2$ pulse to prepare a superposition between $\ket{\uparrow}$ and $\ket{\downarrow}$ for each ion. Ideally all four measurement outcomes should have the same probability of $25\,\%$. The deviation from $100\,\%$ is due to rounding.}
\end{table}

\subsection{Crosstalk}
\label{sec:crosstalk}
\begin{figure}
	\centering
	\includegraphics[]{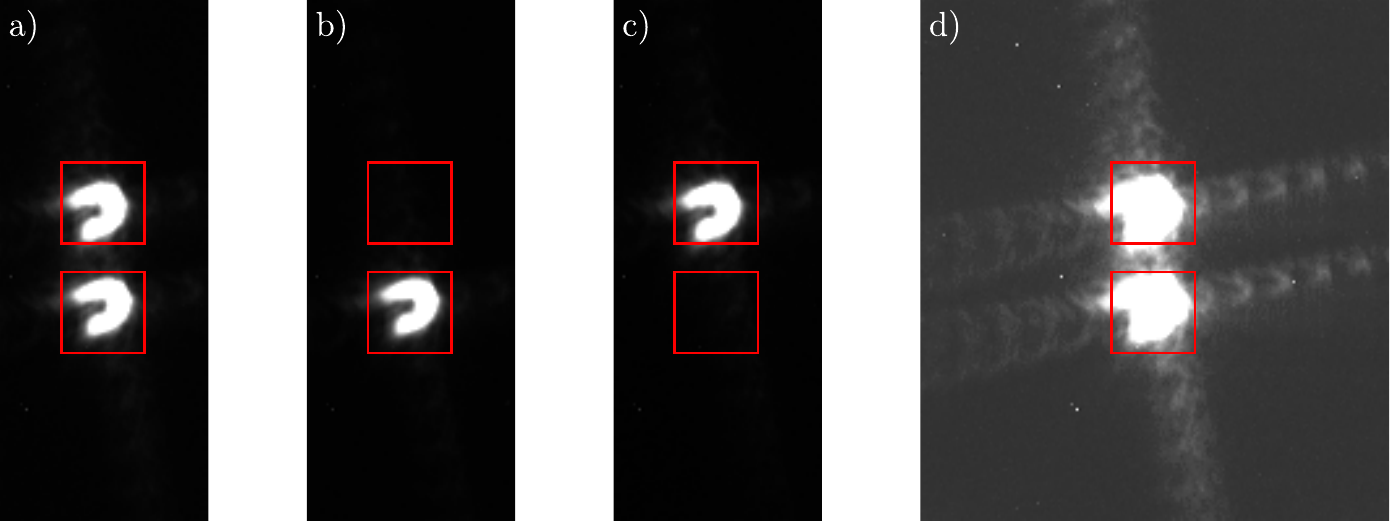}
	\caption{Ion and binning area configuration for crosstalk measurement. Red squares indicate binning areas. a) Two-ion crystal for reference, b) single ion in bottom position, c) single ion in top position.
	d) Same picture as a) but with heavily adjusted contrast to emphasize diffraction spikes.}
	\label{fig:crosstalkImages}
\end{figure}

When reading out the individual qubit states with the camera, one can expect that errors due to crosstalk will increase the SPAM error. Crosstalk means that light emitted from one ion hits a pixel in the binning area of the other ion. To quantify the amount of crosstalk, we first trapped two ions in the trap and took a picture with a long exposure time (100 s) as reference for the next steps; see figure \ref{fig:crosstalkImages}. This measurement was done with two different strenghts of the axial confinement, which changes the center-to-center distance between the ions from $4.6\,\mathrm{\mu m}$ to $5.5\,\mathrm{\mu m}$ to investigate the effect on the amount of crosstalk. We set suitable individual binning areas for the two ions. Since we are currently not able to individually prepare the ions in different states, we cannot investigate the amount of crosstalk with both ions present. Therefore we removed one ion from the trap and applied an electrostatic field on top of the trapping potential to move the remaining ion successively to each of the positions where ions in the two-ion crystal had previously been located. We checked that the position was accurate to within one pixel, which corresponds to approximately $280\,\mathrm{nm}$. For both ion positions and both axial confinement strenghts, we prepared the ion alternating in the bright state and in the dark state before detecting the counts in the two binning areas. The results are shown in table \ref{tab:crosstalk}.

\begin{table}
\centering
\begin{tabular}{c|c|c|c|c|c}\cline{2-5}
	& \multicolumn{2}{|c|}{\textbf{Global bright preparation}} & \multicolumn{2}{|c|}{\textbf{Global dark preparation}} & \\
	\hline
	\multicolumn{1}{|c|}{Ion position} & Area 1 counts & Area 2 counts & Area 1 counts & Area 2 counts & \multicolumn{1}{|c|}{Crosstalk} \\
	\hline
	\multicolumn{6}{|l|}{\textbf{Large separation}}\\
	\hline
	\multicolumn{1}{|c|}{Area 1} & 7764(6) & 1259.7(2.6) & 1196.5(2.7) & 1166.3(2.5) & \multicolumn{1}{|c|}{$1.39(5)\,\%$} \\
	\hline
	\multicolumn{1}{|c|}{Area 2} & 1239.8(2.6) & 7705(6) & 1157.9(2.6) & 1201.7(2.6) & \multicolumn{1}{|c|}{$1.24(6)\,\%$} \\
	\hline
	\multicolumn{6}{|l|}{\textbf{Small separation}}\\
	\hline
	\multicolumn{1}{|c|}{Area 1} & 9833(7) & 1295.9(2.6) & 1168.9(2.7) & 1182.2(2.5) & \multicolumn{1}{|c|}{$1.75(4)\,\%$} \\
	\hline
	\multicolumn{1}{|c|}{Area 2} & 1284.3(2.7) & 9984(7) & 1127.5(2.5) & 1235(2.8) & \multicolumn{1}{|c|}{$1.29(4)\,\%$} \\
	\hline
\end{tabular}

\caption{Results for the crosstalk measurement. The numbers indicate the average number of counts detected in the two binning areas for global preparation in the bright state or the dark state for the two inter-ion distances. Numbers in brackets indicate the standard deviation. Area 1 is the bottom area.}
\label{tab:crosstalk}
\end{table}

To estimate the amount of crosstalk, we compare the amount of light detected in both binning areas during bright preparation and compare it to the amount of detected light for dark preparation. The main assumption is that only a negligible amount of light hits pixels in the wrong binning area when the ion was prepared in the dark state. We will refer to the number of counts detected in the wrong binning area following preparation in the dark state as $N_{dark}$. The amount of crosstalk is then given by the counts above $N_{dark}$ in the wrong area divided by the counts above $N_{dark}$ in all areas. For the larger separation, the detected crosstalk counts when the ion is in location 1 or 2 are 93(4) and 82(4), respectively. For the smaller separation, these numbers increase to 114(4) and 157(4), respectively.

\begin{figure}
	\includegraphics[]{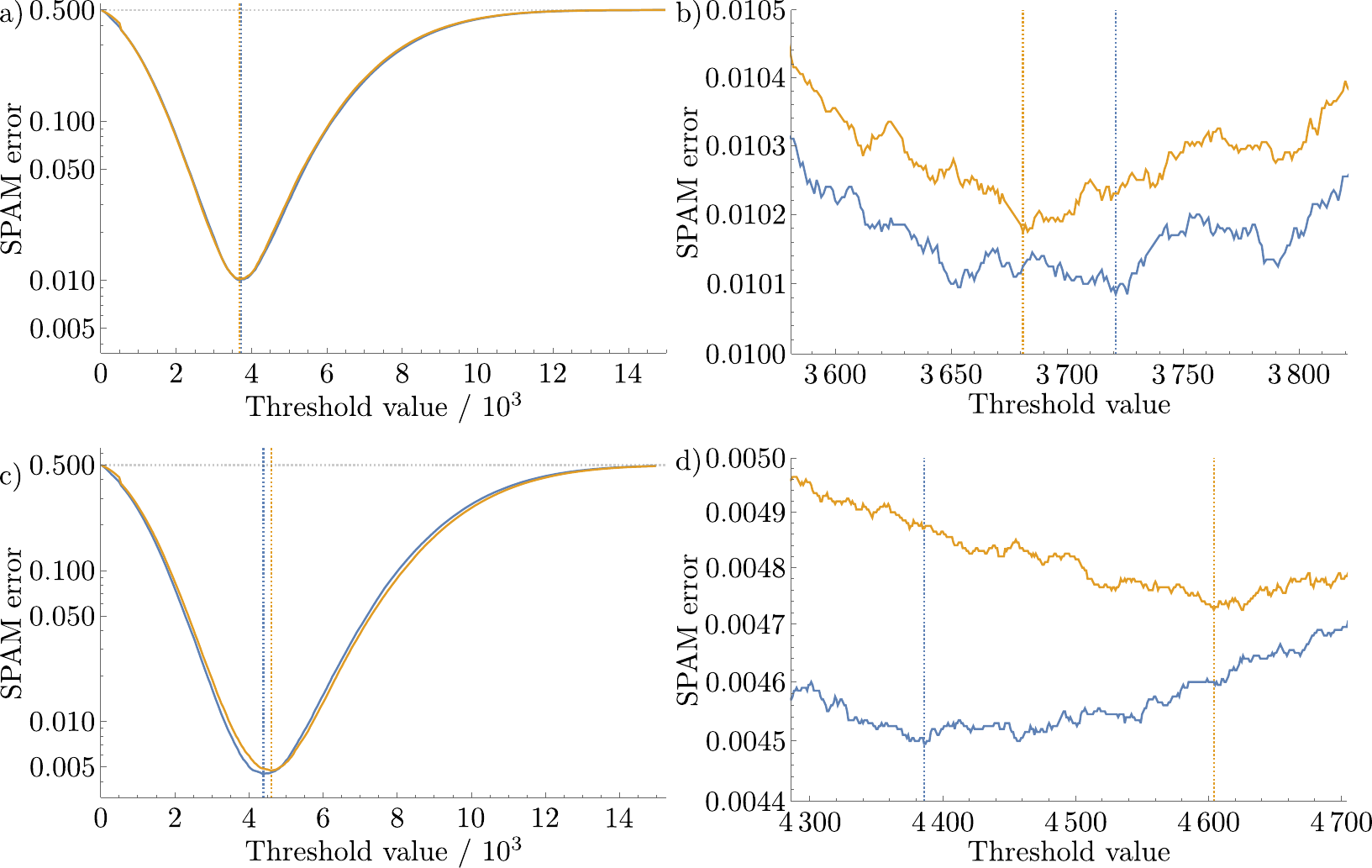}
	\caption{SPAM error as a function of the threshold value for the two ions. 
		First row: larger ion separation, second row: smaller ion separation.
		Blue: first ion, orange: second ion. a,c) Full view; b,d) zoom-in $\pm100$ counts around the threshold values. The vertical colored dashed lines indicate the threshold values for each ion. The horizontal gray dashed line indicates $50\,\%$.}
	\label{fig:thresholderror}
\end{figure}

Here, the crosstalk counts are approximately 100, so if we assume that a neighbouring ion is bright in $50\,\%$ of all detection events, we would adjust the threshold value by half of the crosstalk photons towards higher numbers and the threshold value would be shifted by 50 counts. To judge the effect of 50 counts more or less, we look at how the SPAM error depends on the threshold value (see figure \ref{fig:thresholderror}). The mininum is relatively broad when compared to the amount of extra counts due to crosstalk. One can see that shifts of the threshold value on the order of 100 would not affect the SPAM error significantly. The fluctuations in the data are mainly due to detection noise. The two lines have different minimum values most likely due to different Rabi-rates at different ion positions, which leads to imperfect state transfer to the dark state. Since these fluctuations are one order of magnitude smaller than the minimum SPAM error, we can neglect the effect of crosstalk on the SPAM error in the configuration with the increased distance between the two ions.

When increasing the inter-ion distance, the overall amount of detected light decreases due to the laser power not being compensated after changing the trapping potential, leading to the relative amount of crosstalk being almost constant. This can also be seen in the diffraction spikes in figure \ref{fig:crosstalkImages} d) extending quite far compared to the inter-ion distance. We conclude that the SPAM error is dominated by the amount of detected fluorescence light over the relative amount of crosstalk since the SPAM error is approximately twice as high for the larger inter-ion distance compared to the smaller distance.

\subsubsection{Reasons for crosstalk}
If we assume the ion to be a pointlike source of light, then the image on the CCD sensor is given by the point-spread-function (PSF) of the imaging system, which in our case produces images that look like a backwards 'c' (see figure \ref{fig:crosstalkImages}a). This shape is most likely caused by imperfect alignment of the optical components in the imaging path. In addition we have diffraction spikes in two perpendicular lines crossing the ion's image (see figure \ref{fig:crosstalkImages}b). These spikes are most likely due to diffraction caused by a layer of gold mesh that we use to electrically shield the ions from stray charges on the Schwarzschild objective, which is only $8\,\mathrm{mm}$ away from the ions. The diffraction spikes are in the same line as the two ions and therefore represent our main source of crosstalk. In the future the crosstalk and diffration spikes can be reduced by both rotating the mesh by $45^{\circ}$ so that the diffraction spikes will not coincide with neighboring ion sites, and by using a different mesh size that causes less light to be diffracted.

\section{Detection of coherent oscillations in a multi-qubit system}
\label{sec:twoqubitmultishotperformance}
Now that we have characterized the performance of the PMT and the individual-ion readout capabilities of the camera, we demonstrate the performance of both the camera and the PMT in a two-qubit coherent rotation experiment.

\subsection{Experimental sequence}
We prepare the ions in the $\ket{\uparrow}$ state and drive the qubit transition with resonant microwave pulses of varying lengths. The experiment is repeated 100 times for each pulse duration and the number of counts is averaged over each of the experiments. We perform the experiment with different pulse lengths up to $1300\,\mathrm{\mu s}$ with a step size of $0.4\,\mathrm{\mu s}$.

\subsection{PMT measurement}
\begin{figure}
	\centering
	\includegraphics[]{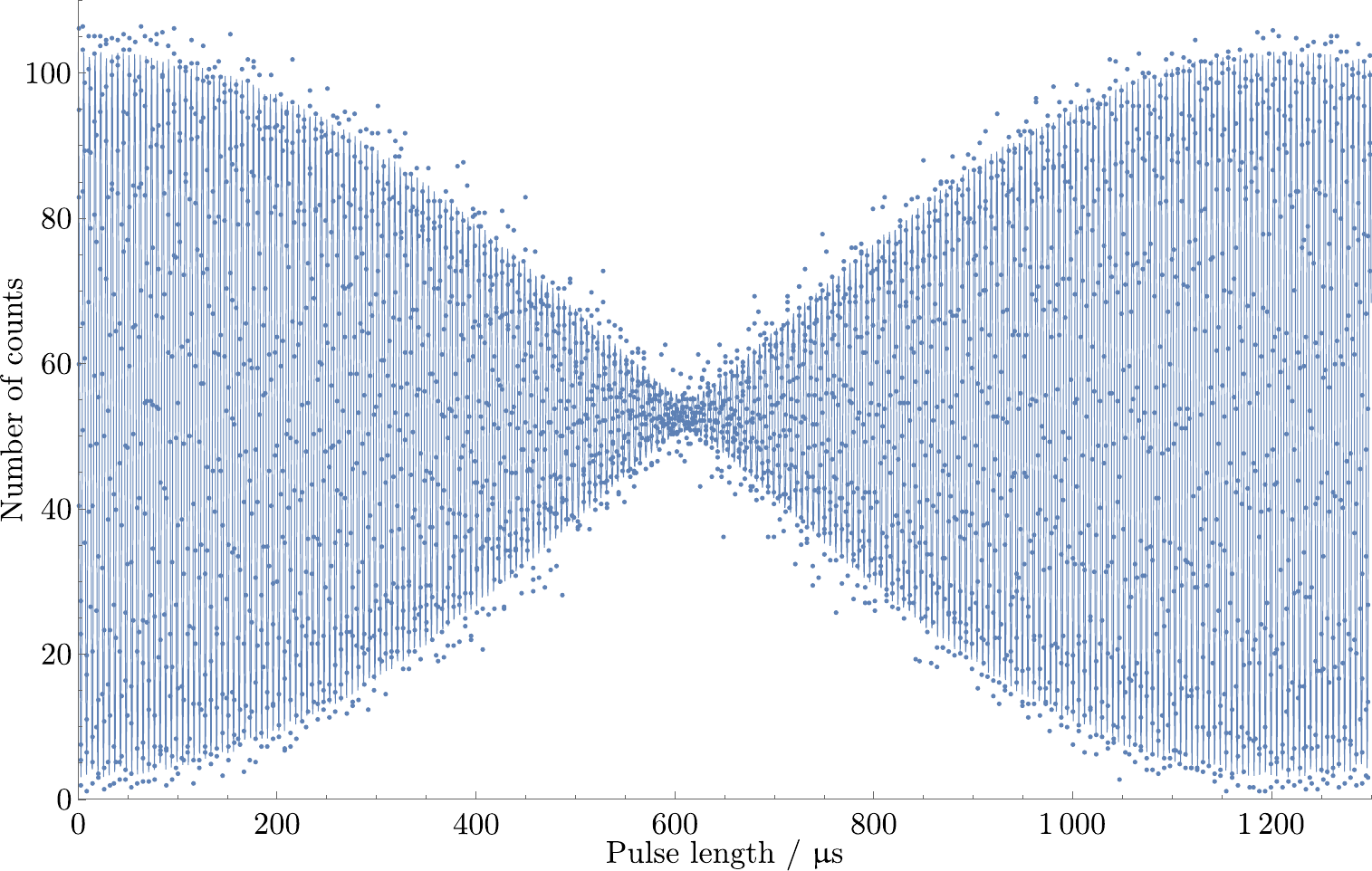}
	\caption{Rabi oscillations with two ions detected with a PMT. Points: experimental data, line: fit. The beat signal is caused by different Rabi frequencies of the two ions.}
	\label{fig:twoionfloppingpmt}
\end{figure}

We observe that the number of counts from both ions decreases with longer pulse durations. After approximately $600\,\mathrm{\mu s}$ the amplitude reaches its minimum value and starts to increase again. This means that the loss in contrast can be traced back to different Rabi frequencies of the two ions which produce a beat signal. The fitted curve shown in figure \ref{fig:twoionfloppingpmt} is of the form $C_{P}(t)=A_1 \cdot \sin(\frac{\pi}{t_{\pi_1}}t+\phi_1) + A_2 \cdot \sin(\frac{\pi}{t_{\pi_2}}t+\phi_2) + B$, where $A_1$ and $A_2$ describe the amplitude of the two Rabi oscillations with an average number of counts $B$. $t_{\pi_i}$ and $\phi_i$ describe the $\pi$-time and phase of the Rabi oscillation of the $i$th ion. The best fit yields $t_{\pi_1}=2.83557(7)\,\mathrm{\mu s}$ and $t_{\pi_2}=2.84904(8)\,\mathrm{\mu s}$, which differ by approximately $0.5\,\%$. Due to the limitation that the PMT can only do a global measurement of the total number of photons, the fit cannot tell which of the two ions has which Rabi frequency.

A possible explanation for the different $\pi$-times is that the two-ion crystal is tilted relative to the microwave conductor which produces the microwave field to drive the qubit transition. If an equal Rabi rate on both ions is desired, this angle could be compensated with suitable DC fields. On the other hand, it is also possible to further enhance this Rabi-frequency difference, possibly also exploiting micromotion sideband transitions~\cite{warring_individual-ion_2013}, to implement individually-addressed single-qubit operations in multi-ion crystals.  This could be very useful in future trapped-ion quantum simulators and quantum computers.

\subsection{Camera detection}
\begin{figure}
	\includegraphics[width=\linewidth]{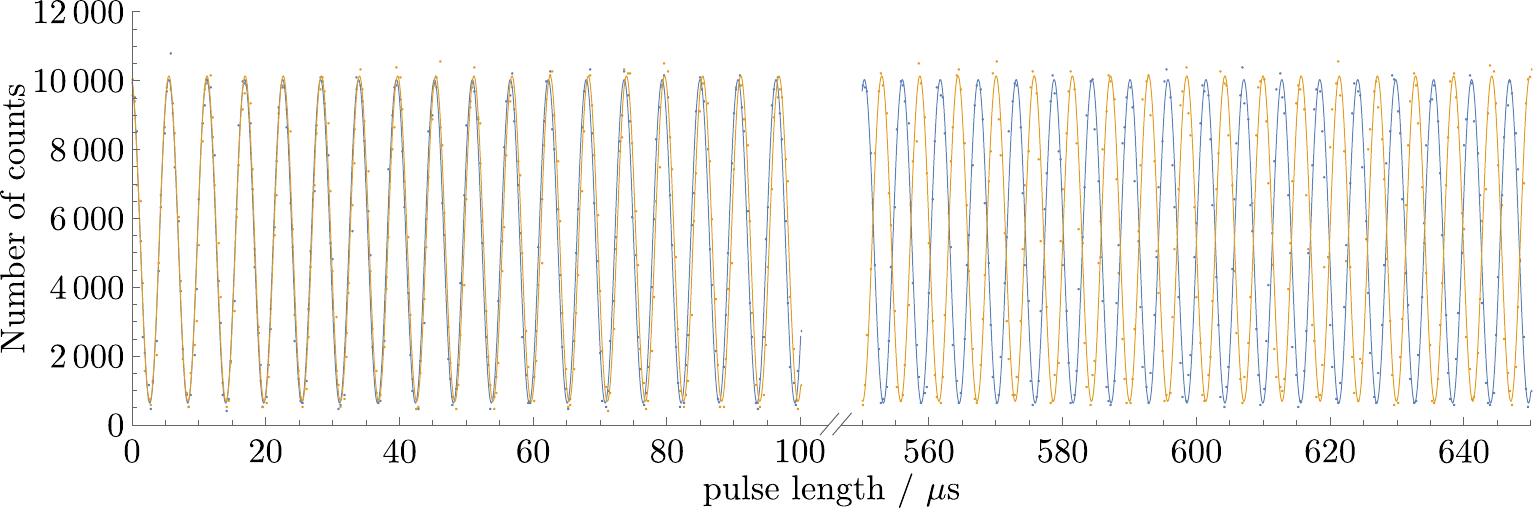}
	\caption{Rabi flopping with two ions detected with an EMCCD camera. Points: experimental data, lines: fit. The oscillation can be detected for each ion individually. Blue: ion 1, Orange: ion 2}
	\label{fig:2ionscamera}
\end{figure}

We then repeated the same experimental sequence while detecting the fluorescence with the EMCCD camera. Each ion is assigned its own binning area, so we are able to follow the oscillation of both ions individually (see figure \ref{fig:2ionscamera}). From the left part of the plot one can already see that the Rabi rate is slightly higher for ion 1. From the two fits of the form $C_{i}(t)=A_i \cdot \sin(\frac{\pi}{t_{\pi_i}}t+\phi_i) + B_i$ ($i\in\{1,2\}$ for the first/second ion), we can determine the $\pi$-times to be $t_{\pi,1}=2.837688(27)\,\mathrm{\mu s}$ and $t_{\pi,2}=2.851138(27)\,\mathrm{\mu s}$. These $\pi$-times are not exactly the same as we found during the PMT detection experiments, since they can drift over time, possibly due to temperature dependence of the microwave amplifiers used to drive the hyperfine transitions. In this specific case the relative drift of the $\pi$-time was around $10^{-4}$ per hour.

\section{Conclusion}
We have demonstrated that the performance of a real-time readout capable UV EMCCD camera is suitable for quantum simulation and quantum computation experiments with trapped $^{9}\mathrm{Be}^{+}$ ions. In contrast to a PMT, the real-time capable camera can determine the qubit state for each individual ion with a negligible effect of crosstalk from neighboring ions on the SPAM error. The fast real-time capable measurement may enable the implementation of quantum error correction protocols. It may also benefit new clock interrogation and fast feedback protocols for quantum metrology applications. 
\medskip

\textbf{Acknowledgements} \par
We acknowledge funding by DFG through SFB 1227 ``DQ-mat" (project A01), the cluster of excellence ``QuantumFrontiers", the European Union through the QT flagship project ``MicroQC" and the Volkswagen foundation and the Ministry of Science and Culture of Lower Saxony through ``Quantum Valley Lower Saxony Q1" (QVLS-Q1).
\medskip

\bibliographystyle{MSP}
\bibliography{Camera_Paper}

\end{document}